\documentclass[11pt,a4paper]{article}
\usepackage{jcappub}

\pdfoutput=1
\usepackage{euscript}
\usepackage{amssymb}
\usepackage{amsfonts}
\usepackage{amsbsy}
\usepackage{amsmath}
\usepackage{epsfig}
\usepackage{amsthm}
\usepackage{amscd}
\usepackage{amstext}
\usepackage{verbatim}

\usepackage{color}

\newcommand{\beq}{\begin{equation}}
\newcommand{\eeq}{\end{equation}}
\newcommand{\bea}{\begin{eqnarray}}
\newcommand{\eea}{\end{eqnarray}}

\title{Out of the White Hole:\\ {\it A Holographic Origin for the Big Bang}}

\author[a,b]{Razieh Pourhasan,}
\author[a,b]{Niayesh Afshordi,}
\author[a,b]{and Robert B. Mann}

\affiliation[a]{Department of Physics \& Astronomy, University of
Waterloo, Waterloo, Ontario N2L 3G1, Canada}
\affiliation[b]{Perimeter Institute for Theoretical Physics, 31
Caroline St. N., Waterloo, ON, N2L 2Y5, Canada}

\emailAdd{r2pourhasan@uwaterloo.ca}
\emailAdd{nafshordi@pitp.ca}
\emailAdd{rbmann@uwaterloo.ca}

\abstract{While most of the singularities of General Relativity
are expected to be safely hidden behind event horizons by the
cosmic censorship conjecture, we happen to live in the causal
future of the classical {\it big bang} singularity, whose
resolution constitutes the active field of early universe
cosmology. Could the big bang be also hidden behind a causal
horizon, making us immune to the decadent impacts of a naked
singularity? We describe a braneworld description of cosmology
with both 4d induced and 5d bulk gravity (otherwise known as
Dvali-Gabadadze-Porati, or DGP model), which exhibits this
feature: The universe emerges as a spherical 3-brane out of the
formation of a 5d Schwarzschild black hole.  In particular, we
show that a pressure singularity of the holographic fluid,
discovered earlier, happens inside the white hole horizon, and
thus need not be real or imply any pathology. Furthermore, we
outline a novel mechanism through which any thermal atmosphere for
the brane, with comoving temperature of $ \sim 20$\% of the 5D
Planck mass can induce   scale-invariant primordial curvature
perturbations on the brane, circumventing the need for a separate
process (such as cosmic inflation) to explain current cosmological
observations. Finally, we note that 5D space-time is
asymptotically flat, and thus potentially allows an S-matrix or
(after minor modifications) AdS/CFT description of the
cosmological big bang.}

 \arxivnumber{1309.xxxx}

\begin{document}
\maketitle \flushbottom

\section{Introduction}

The scientific discipline of Physical Cosmology started as, and
continues to be, an extremely ambitious attempt to summarize the
physics of the entire universe within a handful of cosmological
parameters. However, maybe the most surprising outcome of this
enterprise has been how successful this naive approach has been in
describing cosmological observations that are multiplying at an
accelerating rate. This is exemplified by the spectacular data
recently released by the Planck collaboration \cite{Ade:2013ktc},
and its remarkable agreement with the six-parameter  $\Lambda$CDM
paradigm. However, the experimental success of standard cosmology
is overshadowed by fundamental existential questions: What is Dark
Matter? Why Dark Energy? What is the nature of the Big Bang?

The starting point for this paper was to ask whether a more
satisfactory (or natural) understanding of these mysteries can
come from an alternative description of the geometry. In
particular, could these (seemingly unrelated) phenomena be
manifestations of hidden spatial dimensions, that show up as
``holographic fluid(s)'' in our 4D description?

Motivated by D-branes in 10D string theory, pure phenomenology, or
a combination of the two, one way to describe our four-dimensional
universe is through embedding it in a higher dimensional
spacetime-- with at least one more dimension-- and investigate its
gravitational and/or cosmological properties. This is known as the
``\textit{brane world}'' scenario,  where the brane refers to our
4D universe embedded in a bulk space-time with 5 or more
dimensions, where only gravitational forces dare to venture.
Well-known (and well-studied) examples of such scenarios are the
Randall-Sundrum (RS)  \cite{Randall:1999ee} model, where 4D
gravity is recovered through a compact volume bulk, or the
Dvali-Gabadadze-Porrati (DGP) construction \cite{Collins:2000yb,Dvali:2000hr},
where our 3-brane is equipped with its own induced gravity,
competing with the bulk gravity via the so-called Vainshtein
mechanism \cite{Vainshtein:1972sx}.

Radiation dominated cosmology has been studied in the context of
RS model where FRW metric describing 4D universe emerges as
induced gravity on the brane in 5D AdS/Shchwarzschild background,
e.g. \cite{Savonije:2001nd, Gubser:1999vj}. However in this paper,
we focus on the DGP model, which is defined by the following
action: \beq S_{DGP} \equiv \frac{1}{16\pi G_b}\int_{\rm bulk} d^5
x \sqrt{ - g} R_5 + \frac{1}{8\pi G_b}\int_{\rm brane} d^4 x
\sqrt{-\gamma} K+ \int_{\rm brane} d^4 x \sqrt{-\gamma}
\left(\frac{R_4}{16\pi G_N} + {\cal L}_{\rm matter}\right), \eeq
where $g$ and $\gamma$ are the bulk and brane metrics
respectively, while $K$ and $R_4$ are the mean extrinsic and Ricci
intrinsic curvatures of the brane. $G_b$ and $G_N$ are then
respectively the bulk and brane (i.e. Newton's) gravitational
constants. One may also express the gravitational constants in
terms of the bulk and brane Planck masses: \beq M_4 = (16 \pi
G_N)^{-1/2} \qquad M_5 = (32 \pi G_b)^{-1/3}, \label{Planck_mass}
\eeq which respectively describe the approximate energies at which
the brane and bulk gravitons become strongly coupled. Moreover,
the ratio $r_c \equiv G_b/G_N$ characterizes  the length scale
above where 5D gravity becomes important.

Along with a great deal of attention, these models have received
some criticism. The DGP model includes a de Sitter solution
automatically, which is usually called a self accelerating (SA)
branch. When first proposed, this gave rise to  the hope of a
consistent description of our accelerating universe  without a
cosmological constant. However, it turned out that SA solutions
suffer from ghosts and tachyons \cite{Pilo:2000et, Luty:2003vm,
Nicolis:2004qq, Charmousis:2006pn} as well as some pathological
singularities \cite{Kaloper:2005wa}.  Furthermore, the detailed
predictions of the SA branch were inconsistent with cosmological
observations \cite{Fang:2008kc}.  Nevertheless, the normal
(non-self-accelerating) branch of the DGP cosmology does not
suffer from the same pathologies, and can be consistent with data,
if one includes brane tension (which is the same as a 4D
cosmological constant) \cite{Azizi:2011ys}.

While most studies in the context of DGP have been made from the
viewpoint of a 4D observer living on the brane, the DGP model was
reexamined \cite{Gregory:2007xy}  as a theory of 5D Einstein
gravity coupled to 4D DGP branes,   using a Hamiltonian analysis.
New pathologies were encountered in the model by generalizing the
5D geometry from Minkowski space-time -- as originally considered
in the DGP model-- to Schwarzschild. If the black hole mass in the
bulk exceeds a critical value, a so called ``\textit{pressure
singularity}'' will arise at finite radius \cite{Gregory:2007xy}.
Furthermore, on the SA branch the five-dimensional energy is
unbounded from below.

Here we study the DGP model around a 5D black hole in greater
detail to better understand its phenomenological viability. We
relate bulk, brane, and black hole parameters and investigate
constraints on them that allow one to  avoid the pressure
singularity. We find that viable solutions are indeed possible,
leading us to  propose a holographic description for the big bang,
that avoids the big bang singularity. We further outline a  novel
mechanism through which the brane's atmosphere induces (near)
scale-variant curvature perturbations on the brane, without any
strong fine tuning (or need for additional processes, such as
cosmic inflation), consistent with cosmic microwave background
observations.

In Section \ref{FRW}, we introduce the induced gravity on the
brane by solving the  vacuum Einstein equations while we demand a
Freedman-Robertson-Walker (FRW) metric on the brane. In Section
\ref{holography}, we describe the geometry in the bulk in more
detail  and clarify the holographic picture of the brane from the
point of view of the 5D observer. We then give our proposal for a
holographic big bang as emergence from a collapsing 5D black hole.
Section \ref{atmosphere} outlines a mechanism to generate
cosmological curvature perturbations from thermal fluctuations in
the brane atmosphere. Finally, Section \ref{conclude}
 wraps up the paper with a summary of our results and related discussions.

\section{Universe with FRW metric}
\label{FRW}

We start by introducing the standard form of the FRW line element:
\begin{equation}
ds^2=-d\tau^2+\frac{a^2(\tau)}{{\cal K}}\left[d\psi^2+\sin\psi^2\left(d\theta^2+\sin^2\theta
d\phi^2\right)\right],\label{line}
\end{equation}
where ${\cal K}>0$ is the curvature parameter whose dimensions are
$(length)^{-2}$ and the scale factor $a$ is dimensionless and
normalized to unity at the present time, i.e. $a(\tau_0) \equiv a_0=1$.

Using  the metric (\ref{line}) for the brane, we next turn to
solving the Einstein equations on the brane
\begin{equation}
G_{\mu\nu}=8\pi G_N(T_{\mu\nu}+\widetilde{T}_{\mu\nu})\,,\label{Ein}
\end{equation}
where $G_N$ is the gravitational constant on the brane. We here
include two types of energy-momentum tensor $T_{\mu\nu}$ and
$\widetilde{T}_{\mu\nu}$. The former describes normal matter
living on the brane in a form of a perfect fluid
$$
T_{\mu\nu}=(P+\rho)u_{\mu}u_{\nu}+Pg_{\mu\nu},
$$
satisfying the continuity equation
\begin{equation}
\nabla^{\mu}T_{\mu\nu}=0, \label{conT}
\end{equation}
where $g_{\mu\nu}$ is the metric on the brane given by
(\ref{line}) and $u^{\mu}$ is the 4-velocity of the fluid
normalized such that $u^{\mu}u_{\mu}=-1$. 
The latter stress-energy  $\widetilde{T}_{\mu\nu}$ is the Brown-York stress
tensor \cite{Brown:1992br}  induced on the brane, defined from the extrinsic curvature
$K_{\mu\nu}$ as
\begin{equation}
\widetilde{T}_{\mu\nu}\equiv\frac{1}{8\pi G_b}\left(
Kg_{\mu\nu}-K_{\mu\nu}\right),\,\label{BY}
\end{equation}
where $G_b$ is the gravitational constant in the bulk,   and we
have assumed $Z_2$ bulk boundary conditions on the brane. The
vacuum Einstein equations in the bulk impose  the following
constraints on the brane
\begin{eqnarray}
&&\nabla^{\mu}\left(Kg_{\mu\nu}-K_{\mu\nu}\right)=0\,,\label{conMom}\\
&&R+K^{\mu\nu}K_{\mu\nu}-K^2=0\,,\label{conEn}
\end{eqnarray}
where $R=-8\pi G_N(T+\widetilde{T})$ is the Ricci scalar on the brane.
The first constraint is just the continuity equation for
$\widetilde{T}_{\mu\nu}$ while the second one is the so called
Hamiltonian constraint.

Without loss of generality, as a result of the symmetry of FRW
space-time, we can write $\widetilde{T}_{\mu\nu}$ in a perfect
fluid form i.e.
\begin{equation}\label{braneflu}
\widetilde{T}_{\mu\nu}=(\widetilde{P}+\widetilde{\rho})u_{\mu}u_{\nu}+\widetilde{P}g_{\mu\nu},
\end{equation}
which we shall refer to as the induced (or holographic) fluid.
Combining Eqs. (\ref{BY}) and (\ref{braneflu}), we get: \beq
K_{\mu\nu} = -8\pi G_b\left[
(\widetilde{P}+\widetilde{\rho})u_{\mu}u_{\nu}+\frac{1}{3}
\widetilde{\rho}g_{\mu\nu} \right].\label{k_rho_p} \eeq
From Eqs.
(\ref{Ein}), (\ref{conT}), (\ref{conMom}) and (\ref{conEn}) we
respectively obtain
\begin{eqnarray}
&&H^2+\frac{{\cal K}}{a^2}=\frac{8\pi G_N}{3}(\rho+\widetilde{\rho})\,,\label{E1}\\
&&\dot{\rho}+3H(\rho+P)=0\,,\label{E2}\\
&&\dot{\widetilde{\rho}}+3H(\widetilde{\rho}+\widetilde{P})=0\,,\label{E3}\\
&&\widetilde{\rho}+\rho-3(P+\widetilde{P})+\frac{8\pi G_b^2}{G_N}\left(\frac{2}{3}\widetilde{\rho}^2+2\widetilde{\rho}\widetilde{P}\right)=0, \label{E4}
\end{eqnarray}
where the last equation follows from solving for $K_{\mu\nu}$ in
terms of $(\widetilde{\rho},\widetilde{P})$ using Eq. (\ref{k_rho_p}).

Combining (\ref{E1}-\ref{E4}) we get for
$\widetilde{\rho}$ and $\widetilde{P}$:
\begin{eqnarray}
&&\widetilde{\rho}_{\pm}=\widetilde{\rho}_s\left(1\pm
\sqrt{1-\frac{\mu^2}{12\pi G_N\widetilde{\rho}_s}\frac{1}{a^4}+\frac{2\rho}{\widetilde{\rho}_s}}\right)\,,\label{rhot}\\
&&\widetilde{P}=\frac{\widetilde{\rho
}^2+\widetilde{\rho}_s(\widetilde{\rho
}-T)}{3(\widetilde{\rho}_s-\widetilde{\rho }) }\, ,\label{pt}
\end{eqnarray}
where
\begin{equation}
T=3P-\rho\,
\end{equation}
 and we choose the constant of integration  $-\mu^2$, of dimension $[{ \rm length}]^{-2}$, to be negative
(see e.g. \cite{Maeda:2003ar}, for a similar derivation of DGP cosmology).
The choice of
minus sign will be justified in the next section, where we
introduce the holographic picture.  Finally, we have also defined the characteristic density scale for the holographic fluid:
\begin{equation}
\widetilde{\rho}_s\equiv\frac{3G_N}{16\pi G_b^2}\,.
\end{equation}

Equation (\ref{pt}) immediately implies that the pressure becomes singular
at $\widetilde{\rho}=\widetilde{\rho}_s$.  It is then of interest to investigate
in whether this pressure
singularity can happen at early or late times (if at all), in our cosmic history.
 We  address this
question in the next section.

Furthermore, we note that $\tilde{\rho}_s$ sets the characteristic
density scale, below which the bulk gravity becomes important.
Specifically, it is easy to see that both terms in the induced
fluid density, $\tilde{\rho}$ (Eq. \ref{rhot}), become much
smaller than the matter density, $\rho$, if $\rho \gg
\tilde{\rho}_s$. Therefore, given the current lack of
observational evidence for 5D gravity (e.g. \cite{Azizi:2011ys}),
it is safe to assume that $\rho(z) > \rho_{\rm now} \gg
\tilde{\rho}_s$, i.e. the induced fluid has always had a
negligible contribution to cosmic expansion, with the notable
(possible) exception of the above-mentioned singularity.

\section{Universe as a hologram for a Schwarzschild bulk}
\label{holography}

Consider  our universe to be a (3+1)-dimensional holographic
image \cite{Susskind:1994vu} -- call it a brane -- of a
(4+1)-dimensional background Schwarzchild geometry
\begin{equation}
ds^2_{\mathrm{bulk}}=-f(r)dt^2+\frac{dr^2}{f(r)}+r^2d\Omega_3^2\,,\label{bulkmetric}
\end{equation}
with
\begin{equation}
f(r)=1-\frac{r_h^2}{r^2}\,,
\end{equation}
and where $d\Omega_3$ is the metric of unit 3-sphere. We now
assume a dynamical brane, i.e our universe, to be located at
$r=a(\tau)/\sqrt{\cal K}$ described by the FRW metric (\ref{line}). Its
unit normal vector is
\begin{equation}
n^{\alpha}=\varepsilon\left(\frac{\dot{a}}{\sqrt{\cal K}f(a)},\sqrt{f(a)+\frac{\dot{a}^2}{\cal K}},0,0,0\right)\,,
\end{equation}
with $n^{\alpha}n_{\alpha}=1$ and $\varepsilon=-1$ or $+1$, and we
take
\begin{equation}
u^{\alpha}=\left(\frac{1}{f(a)}\sqrt{f(a)+\frac{\dot{a}^2}{\cal K}},\frac{\dot{a}}{\sqrt{\cal K}},0,0,0\right)
\end{equation}
 to be the unit timelike tangent vector on the brane, i.e
$u^{\alpha}u_{\alpha}=-1$.

Recall that  besides normal matter on the brane we also introduced
an induced fluid denoted by $\widetilde{T}_{\mu\nu}$ on the brane,
which is the imprint of the bulk geometry through the junction
condition (\ref{BY}). Using
\begin{equation}
K_{ab}=n_{\alpha;\beta}e_{a}^{\alpha}e_{b}^{\beta}
\end{equation}
with $a, b$ and $\alpha, \beta$ labelling the brane and bulk
coordinates  respectively, it is just a matter of calculation to
obtain
\begin{eqnarray}
&&K_{ij}=\frac{\varepsilon\sqrt{{\cal K}}}{a}\sqrt{f(a)+\frac{\dot{a}}{\sqrt{{\cal K}}}}\,\Omega_{ij}\,,\label{kij}\\
&&K_{\tau\tau}=-\frac{\varepsilon(\frac{{\cal K}^2
r_h^2}{a^4}+\frac{\ddot{a}}{a})}{\sqrt{H^2+\frac{{\cal K}}{a^2}-\frac{{\cal K}^2
r_h^2}{a^4}}}\,,\label{ktt}
\end{eqnarray}
where $\tau$ is the proper time of the brane and $i, j$ label the
coordinates of the spatial section, with $H\equiv\dot{a}/a$ is the
Hubble parameter. $\Omega_{ij}$ is the metric of the unit 3-sphere. Using (\ref{kij}-\ref{ktt}) for the extrinsic
curvature in (\ref{BY}) and considering $\widetilde{T}_{\mu\nu}$ in a
form of a perfect fluid on the brane we find
\begin{eqnarray}
&&\widetilde{\rho}_{\pm}=\widetilde{\rho}_s\left(1\pm
\sqrt{1-\frac{2(\rho_{_{\rm BH}}-\rho)}{\widetilde{\rho}_s}}\right)\,,\label{rhotilde}\\
&&\widetilde{P}=\frac{-\left(1+2\varepsilon \right)\widetilde{\rho
}^2+\widetilde{\rho}_s(\widetilde{\rho
}-T)}{3(\widetilde{\rho}_s+\varepsilon \widetilde{\rho })
}\, ,\label{Ptilde}
\end{eqnarray}
where $\rho_{_{\rm BH}}$ is a characteristic 3-density,
proportional to the density of the bulk black hole, averaged
within our 3-brane, defined as:
\begin{equation}
\rho_{_{\rm BH}} \equiv \frac{3\Omega_k^2 H_0^4 r_h^2}{8\pi G_N a^4}
\, ,\label{rhoc}
\end{equation}
while $\Omega_k\equiv-{\cal K}/H_0^2$. Comparing (\ref{rhot}) with
(\ref{rhotilde}) we see that the integration constant $\mu$ from
the previous section could be interpreted as the mass of the black
hole in the bulk, given in terms of the horizon radius as
\begin{equation}
\mu=3 |\Omega_k| H_0^2 r_h\,,
\end{equation}
with the comparison between (\ref{Ptilde}) and (\ref{pt}) further
indicating that $\varepsilon=-1$, and as a result, at
$\widetilde{\rho}=\widetilde{\rho}_s$ the pressure becomes
singular.  Moreover, as promised in the previous section, $- \mu^2 \propto -r^2_h < 0$,
which is necessary for positive energy (or ADM mass) initial conditions.

We note that $\widetilde{\rho}_+$ is non-zero, even for
$\rho=\rho_{_{\rm BH}}=0$, which is often known as the {\it
self-accelerating }(SA) branch in the literature, as the universe
can have acceleration, even in the absence of a cosmological
constant (or brane tension). However, as discussed in the
introduction, SA branch suffers from a negative energy ghost
instability. On the other hand, $\widetilde{\rho}_-$, known as the
{\it normal} branch, does not suffer from the same problems, and
may well provide a healthy effective description of bulk gravity
(e.g. \cite{Nicolis:2004qq}). In what follows, we outline
constraints on both branches for the sake of completeness.

In total, we have  three adjustable parameters in our model:
$\widetilde{\rho}_s$, ${\cal K}$, and $r_h$. We shall next
consider the  constraints on these parameters. We find two limits
on $\widetilde{\rho}_s$. One is from demanding reality of all
quantities in (\ref{rhotilde}), i.e.
\begin{equation}
\widetilde{\rho}_s\geq2\left(\rho_{_{\rm BH}}-\rho\right)\,.\label{realitycons}
\end{equation}
where the equality indicates a pressure singularity. The other
comes from the fact that, thus far,  cosmological observations
have not detected any effect of the induced fluid
$\widetilde{\rho}$, which implies that the density of the induced
matter on the brane should be small compared to normal matter in
the universe. These constraints are often expressed in terms of
the transition scale $r_c$ \cite{Deffayet:2000uy}, where
\begin{equation}
r_c \equiv \left(3 \over 16 \pi G_N \tilde{\rho}_s \right)^{1/2}  = \frac{G_b}{G_N},
\end{equation}
which is constrained to be bigger than today's cosmological
horizon scale (e.g., \cite{Azizi:2011ys}). Therefore,  we impose a
conservative bound
\begin{equation}
\mid\widetilde{\rho}\mid\lesssim \epsilon\rho\,,\label{rhocons}
\end{equation}
where $\epsilon\ll1$.

The constraints
(\ref{realitycons}) and  (\ref{rhocons}) restrict the parameter
space. To clarify this we employ equation (\ref{rhotilde}),
investigating the positive and negative branches separately.
\begin{figure}[t]
\centering{\includegraphics[width=300pt,height=200pt]{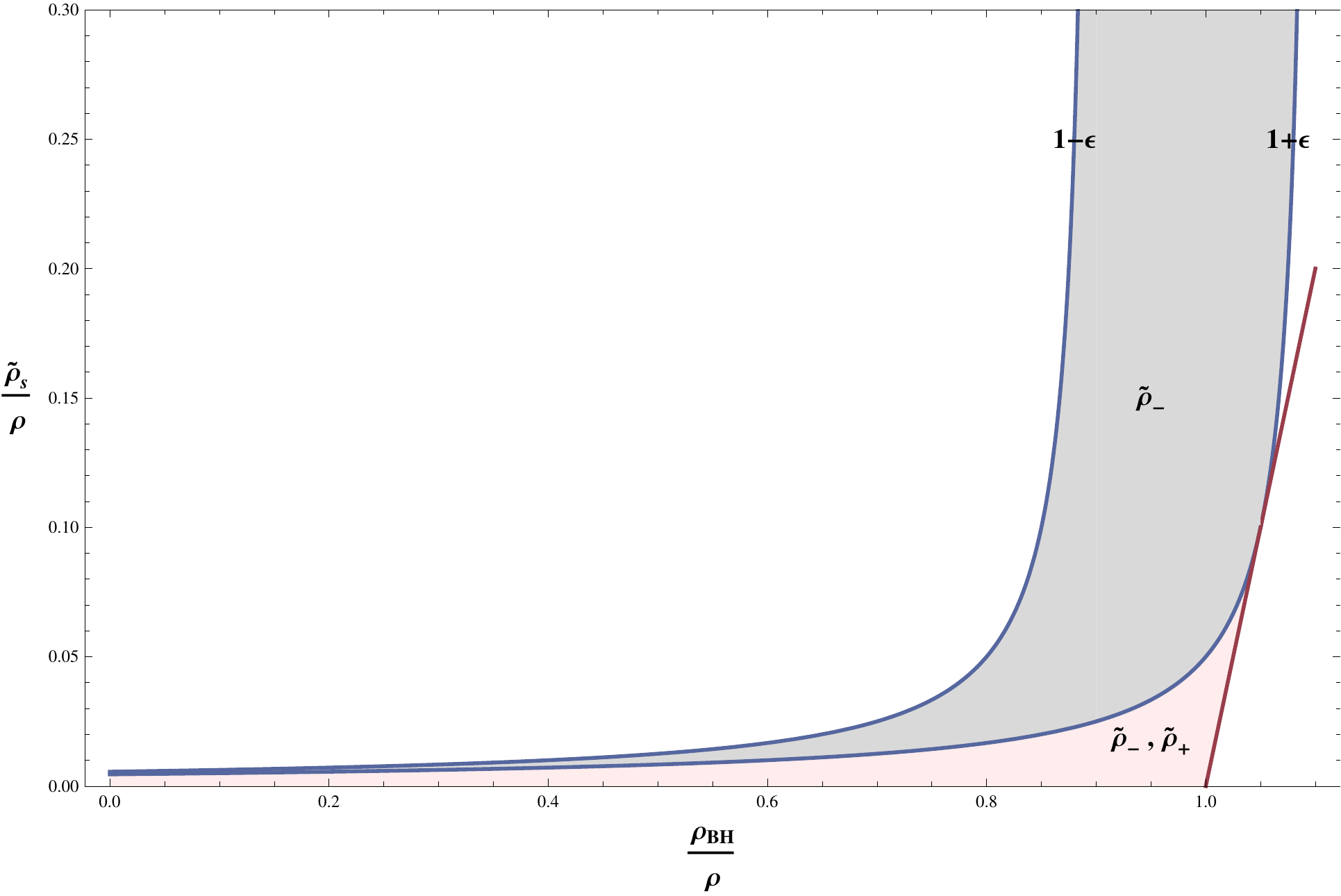}}
\caption{The shaded area shows the allowed values of
$\widetilde{\rho}_s$ and $\rho_{_{\rm BH}}$ for both branches (pink), and only
$\widetilde{\rho}_-$ or the normal branch (gray). The red solid line indicates those
values of $\widetilde{\rho}_s$ and $\rho_{_{\rm BH}}$ for which pressure
becomes singular. We have chosen $|\widetilde{\rho}/\rho| < \epsilon=0.1$ in this
figure.  }\label{fig:rhosvsrhoc}
\end{figure}
Consider first the positive branch. Solving (\ref{rhocons}) for
$\widetilde{\rho}_+$ yields the upper bound
\begin{equation}
\frac{\widetilde{\rho}_s}{\rho}\leq
\frac{\epsilon^2}{2}\left(1+\epsilon-\frac{\rho_{_{\rm BH}}}{\rho}\right)^{-1}, {\rm~for~} \widetilde{\rho}_+ {\rm ~(self-accelearting~branch)},
\,\label{rhosmax}
\end{equation}
which along with eq. (\ref{realitycons}) bounds
$\widetilde{\rho}_s$ within a certain range, i.e. the pink shaded area
in figure (\ref{fig:rhosvsrhoc}). The red line in this figure
shows the values for which pressure becomes singular. Note that
the lower bound (\ref{realitycons}) becomes important only if
$\rho_{_{\rm BH}}>\rho$; condition (\ref{realitycons}) is automatically
satisfied for $\rho_{_{\rm BH}}<\rho$, since $\widetilde{\rho}_s$ is always
positive  by definition. Both upper and lower limits
coincide at $\rho_{_{\rm BH}}=(1+\epsilon/2)\rho$; that is there are upper
bounds for both $\rho_{_{\rm BH}}\leq(1+\epsilon/2)\rho$ and
$\tilde{\rho}_s\leq\epsilon\rho$.

Considering now the negative branch $\widetilde{\rho}_-$ in
(\ref{rhocons}), we obtain upper and lower limits on
$\widetilde{\rho}_s$ as
\bea
\frac{\widetilde{\rho}_s}{\rho}\leq\frac{\epsilon^2}{2}\left(1-\epsilon-\frac{\rho_{_{\rm BH}}}{\rho}\right)^{-1},
{\rm~for~} \frac{\rho_{_{\rm BH}}}{\rho}<1-\epsilon ~(\widetilde{\rho}_-, {\rm ~  normal~branch)}, \label{rhosmaxM1}\\
\frac{\widetilde{\rho}_s}{\rho}\geq\frac{\epsilon^2}{2}\left(1+\epsilon-\frac{\rho_{_{\rm
BH}}}{\rho}\right)^{-1}, {\rm~for~} \frac{\rho_{_{\rm
BH}}}{\rho}>1+\frac{\epsilon}{2}~ (\widetilde{\rho}_-, {\rm
~normal~branch)}.\label{rhosmaxM2} \eea
This allowed region is
shown in Figure (\ref{fig:rhosvsrhoc}) with gray and pink shaded
areas. Note that the red solid line representing the pressure
singularity   sets the lower bound for $\widetilde{\rho}_s/\rho$
within $(1-\epsilon/2)\rho<\rho_{_{\rm BH}}<(1+\epsilon/2)\rho$.

So far we have found limits on $\widetilde{\rho}_s$ and
$\rho_{_{\rm BH}}$. Since the value of $\rho_{_{\rm BH}}$ depends
on the pair $\{\Omega_k, r_h\}$, it is interesting to consider
possible limits on these parameters, and how they affect the
cosmological evolution of our brane. This has been shown in a 3D
plot in Figure \ref{fig:plot3d}. Note that
 any given value for $\rho_{_{\rm BH}}$ in Figure
\ref{fig:rhosvsrhoc} corresponds to a line in the $\{\Omega_k,r_h\}$
plane in Figure \ref{fig:plot3d}. Let us examine this figure more carefully:
\begin{figure}[t]
\centering
{\includegraphics[width=400pt,height=200pt]{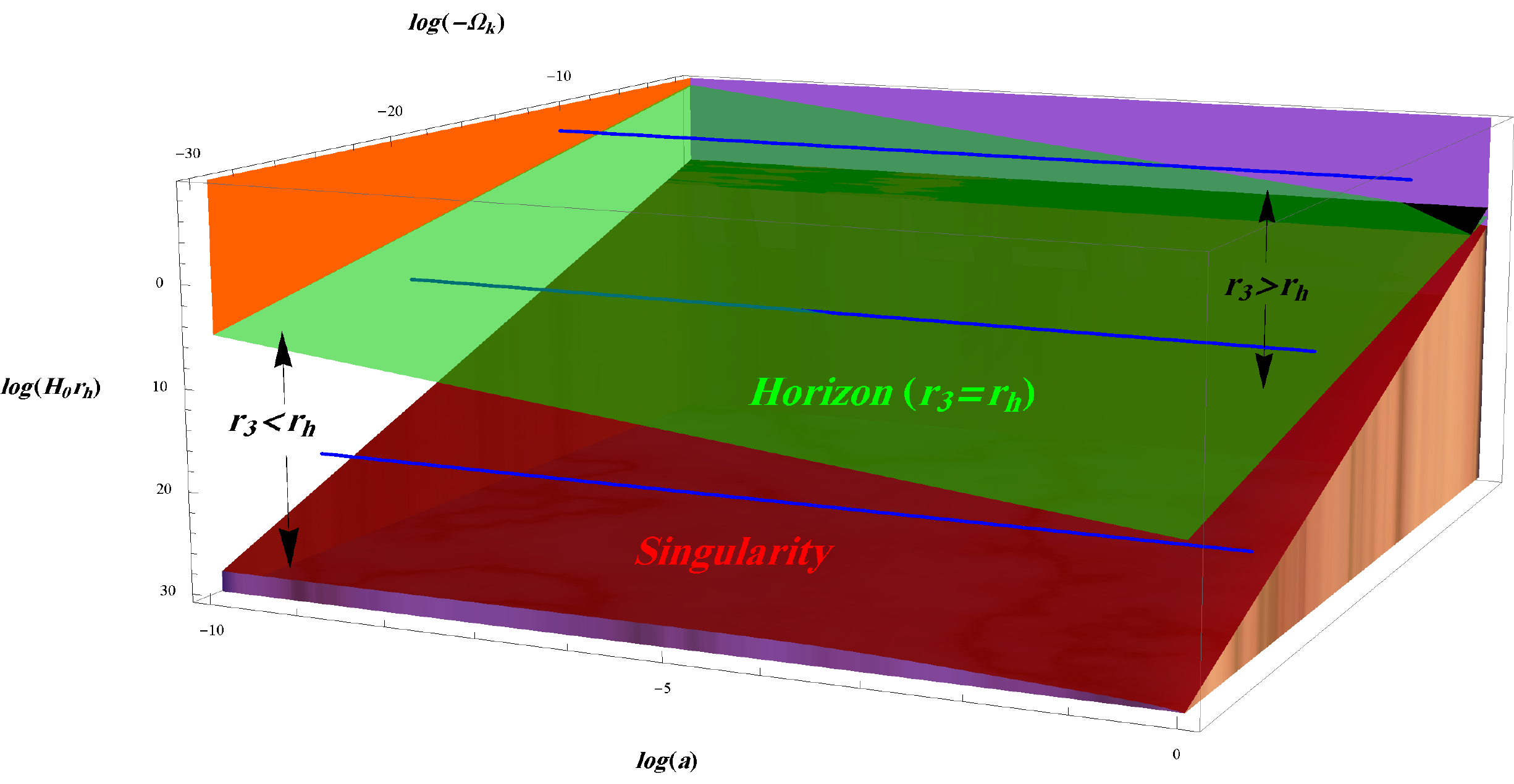}}
\caption{3D plot for $-\Omega_k\leq 0.01$ versus $\log r_h$ from
present time ($a=1$) back to Big Bang Nucleosynthesis ($a\sim
10^{-10}$). The red plane indicates pressure singularity while the
green plane is where $r_h=r_3=a/\sqrt{k}$, i.e. when our brane
leaves the white hole horizon.  The blue lines and the black strip
(visible at the upper right as a triangle, and continuing
underneath the green surface) dicate for a given $\{-\Omega_k,\,
r_h\}$ how the radius of holographic universe evolves from BBN up
to present time; e.g. the black strip represents a holographic
universe that emerges from the pressure singularity during the
radiation era, passes through the white hole horizon at $a\sim
0.01-1$, and eventually is just outside the horizon at the present
time. }\label{fig:plot3d}
\end{figure}

First, note that the figure is plotted for the negative (or normal) branch, which, as discussed above, is physically more relevant. 
The empirical upper limit for the spatial curvature of the
universe $-\Omega_k \lesssim 0.01$ (e.g. \cite{Ade:2013zuv}) is
indicated by the purple vertical plane in the figure. The red
surface represents those pairs of $\{\Omega_k, r_h\}$ for which
$\rho_{_{\rm BH}}=1.05\rho(a)$ from present time ($\log a=0$) back
to Big Bang Nucleosynthesis (BBN; $a \sim10^{-10}$). Here we have
chosen the empirical bound $\epsilon \sim 0.1$ in (\ref{rhocons}),
and have taken BBN as the earliest constraint on deviations from
the standard cosmological model. As we noted before, according to
the reality constraint (\ref{realitycons}) this is the maximum
allowed value for $\rho_{_{\rm BH}}$ at a given time. Therefore
the whole area under the red surface is not allowed. Moreover, the
red surface also shows the possible choices of the pairs
$\{\Omega_k,\,r_h\}$ for which the pressure becomes singular for a
given $a$. Consequently no pressure singularity could happen for
pairs $\{\Omega_k,\,r_h\}$ chosen  to be above the red plane at
any given time.

The green plane indicates those pairs $\{\Omega_k,\,r_h\}$ for
which the radius of our 4 dimensional universe coincides with the
black hole horizon in the 5 dimensional bulk, i.e.
$r_h=r_3=a/\sqrt{k}$.  Therefore, for any $\{\Omega_k,\,r_h\}$
under the green plane, we have  $r_3<r_h$. For those pairs
$\{\Omega_k,\,r_h\}$ chosen to be above  this plane  the radius of
our holographic universe is larger than the horizon radius,
meaning that our present cosmos lies outside the horizon of the
black hole in the bulk, i.e. $r_3>r_h$.  Subsequently, suppose we
choose any pair of $\{\Omega_k,\,r_h\}$ above the green plane at
the present time (the $\log a=0$ plane) and move backwards in
time.  Let us assume that the universe today has its radius larger
than the horizon in the bulk black hole. Moving backwards to early
times ($\log a=-10$ plane),  as the radius of the universe
(proportional to scale factor $a$) decreases, it may or may not
cross the green plane. This has been illustrated with the  upper
two  blue lines in Figure \ref{fig:plot3d}, the lower of which
pierces the green plane at some value of $r_h$ near $\log a \sim
-5$.

Indeed, crossing the green plane means that at some early time the
radius of the universe was smaller than the horizon radius. Since
nothing can escape the horizon of a black hole, one would exclude
those pairs of $\{\Omega_k,\,r_h\}$ for which their  corresponding
blue lines at some $a>10^{-10}$ cross the green plane.
Consequently the pairs highlighted with orange plane are possible
choices of parameters $\{\Omega_k,\,r_h\}$ that satisfy $r_3\geq
r_h$ for $-\Omega_k\leq0.01$ at $a=10^{-10}$.

\begin{figure}[t]
\centering{\includegraphics[width=130pt]{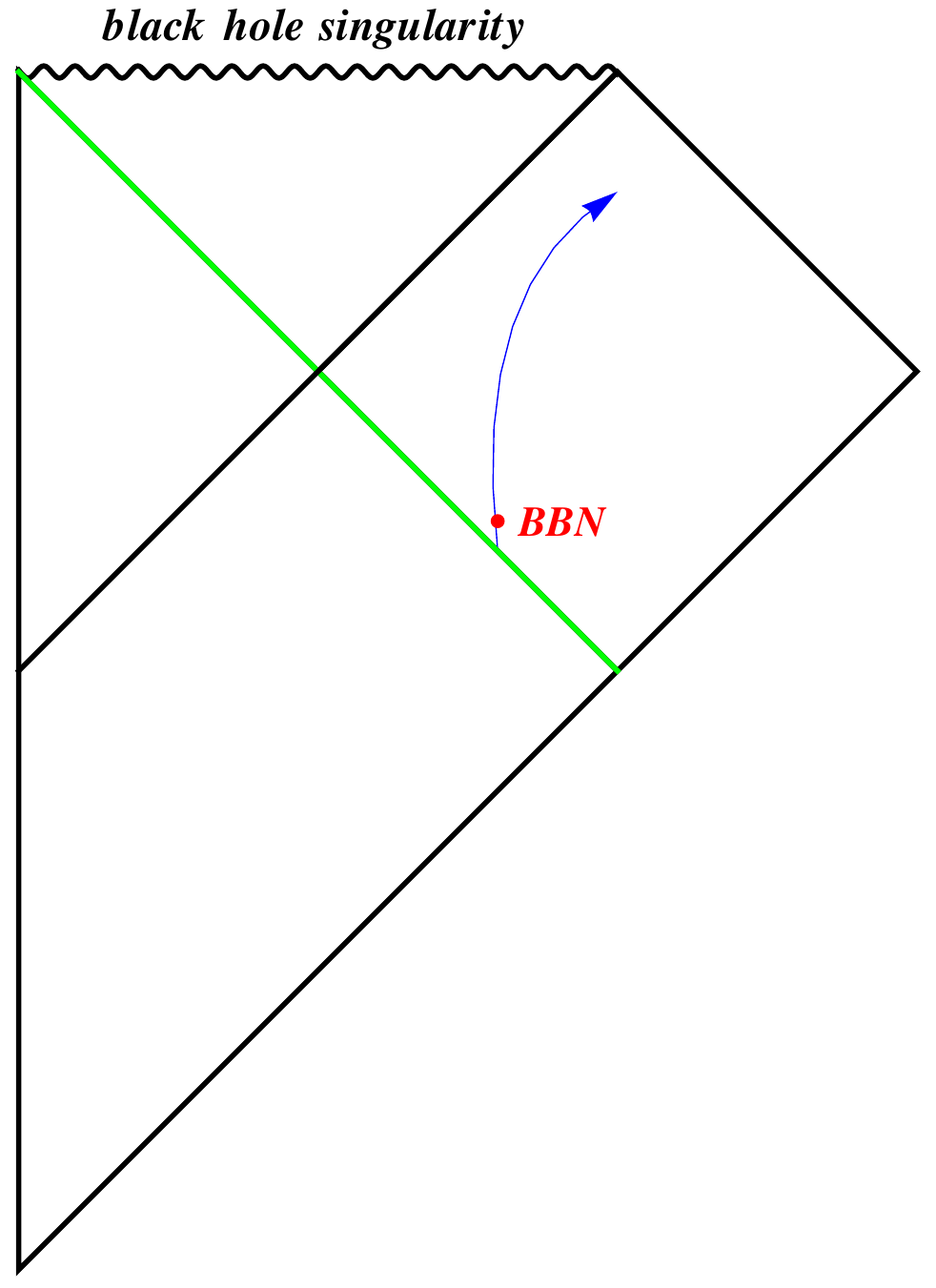}\hspace{40pt}}
\centering{\includegraphics[width=170pt]{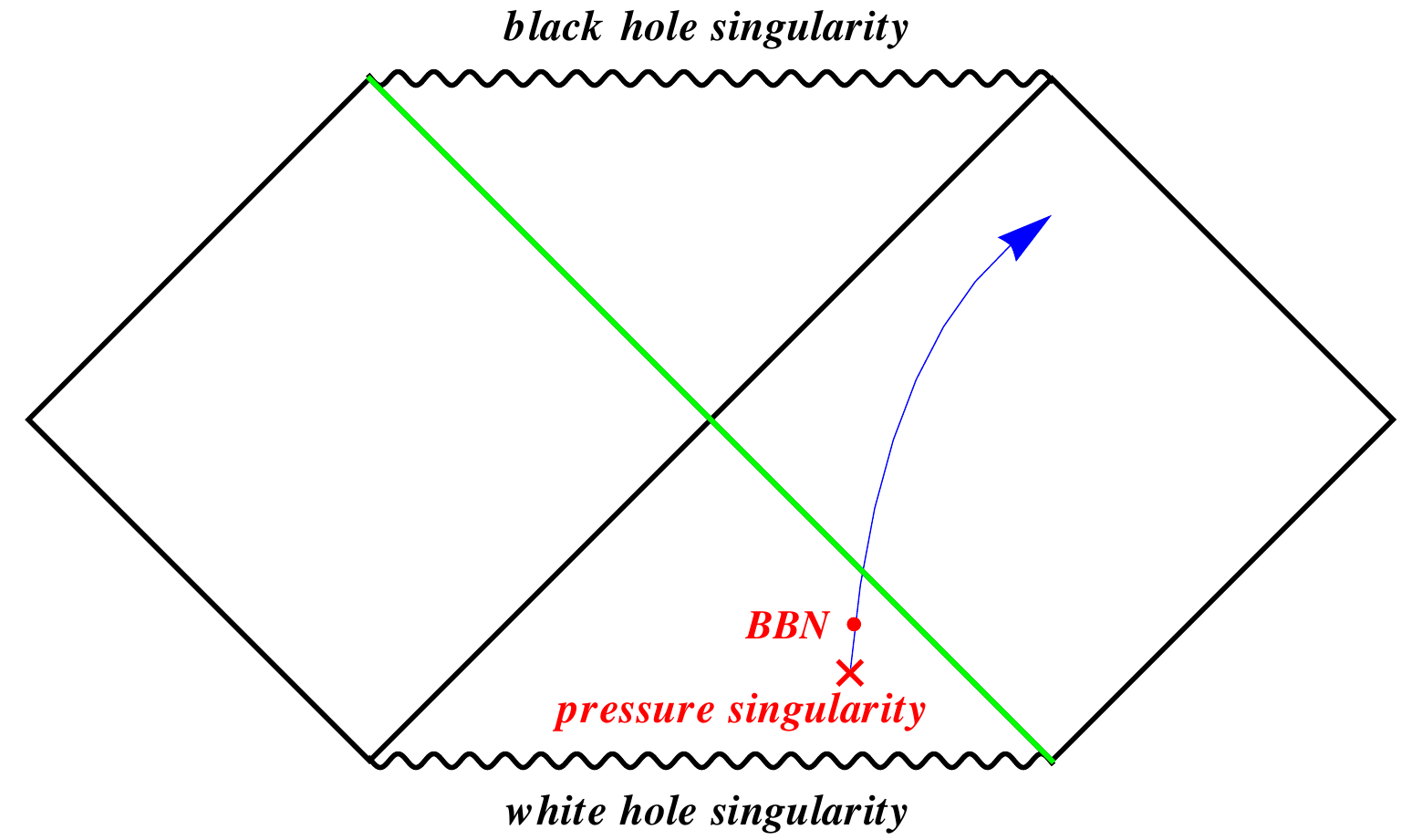}}
\caption{Penrose diagram for the dynamic brane (our universe) in
blue for the black hole  (left) or the white hole  (right) in the
bulk, where the green line indicates  a collapsing shell (or
``star''), or the white hole horizon respectively.
}\label{fig:penrose}
\end{figure}

Consequently,  one may interpret the crossing  $r_3=r_h$ before
BBN ($0<a<10^{-10}$) as the emergence of the holographic universe out of a ``collapsing star'': this scenario replaces the Big Bang singularity. 
The overall picture of this proposal is shown in the
Penrose diagram in Figure (\ref{fig:penrose})-left, which is reminiscent of the core-collapse of a supernova.

Another possibility is to consider a white hole in the bulk rather
than a black hole. With this scenario, it is possible for the
universe to be inside the  horizon at any time up to the present
since all matter eventually emerges from the white hole horizon.
Therefore the entire range of pairs $\{\Omega_k, r_h\}$ above the
red surface is allowed; the lowest blue line in Figure
\ref{fig:plot3d} illustrates one such possible scenario. In this
picture one may interpret the pressure singularity as a
holographic description of the Big Bang that takes place at $a <
10^{-10}$.  Hence those pairs   $\{\Omega_k, r_h\}$ with
$-\Omega_k\leq0.01$  satisfying $\rho_{_{\rm BH}} \lesssim
\rho_r(a=10^{-10})$, i.e. lie above the intersection of the red
surface and the $a=10^{-10}$ plane are allowed\footnote{We have
chosen $\epsilon=0.1$ in (\ref{rhocons}) and
$\rho_r=\rho_0\Omega_r/a^4$.}.  For instance, choosing any value
for $-\Omega_k$ in the range $10^{-4}\leq-\Omega_k\leq10^{-2}$
with its corresponding horizon radius, i.e. $r_h \simeq
\sqrt{\Omega_r}/H_0\Omega_k$, represents a  holographic
universe that emerges from the pressure singularity during the radiation era, 
passes through the white hole horizon at $a\sim 0.01-1$, and
eventually is just outside the horizon at the present time. This
is illustrated with a black strip in Figure \ref{fig:plot3d},
visible at the upper right of the diagram and continuing
underneath the green surface toward the upper left. For any
$-\Omega_k<10^{-4}$, the universe is   inside the horizon at the
present time but (given that its expansion is now dominated by the
cosmological constant), it will expand indefinitely and eventually
intersect the horizon in the future. The overall picture for this
scenario has been shown in the Penrose diagram in Figure
(\ref{fig:penrose})-right.

From the physical point of view, the former scenario, which we can
dub the ``black hole'' universe is more plausible than the latter
``white hole'' universe. The reason is that the region inside a
white hole horizon is to the future of a 4D white hole naked
singularity (Figure \ref{fig:penrose}-right), which makes the
brane dynamics, at best contrived, and at worst ill-defined. In
particular, it is hard to physically justify why this singularity
(i.e. high curvature region) is preceded by a smooth ``zero
temperature'' space-time. For example, it would be in contrast to
(and thus more contrived than) the thermal bath that is the
outcome of the big bang singularity.

\section{Brane Atmosphere and Cosmological  Perturbations}\label{atmosphere}

In this section, we introduce a mechanism to generate
scale-invariant cosmological perturbations in our holographic big
bang. As the holographic fluid is sub-dominant for most of the
cosmic evolution, one expects the standard cosmological
perturbation  theory, that has been extremely successful in
explaining cosmic microwave background observations (e.g.
\cite{Ade:2013ktc,Ade:2013zuv}, amongst other observational
probes), to be applicable. The fluid will dominate  cosmic
evolution at very late times, but that can be avoided for
sufficiently large $r_c$ or small $a$.

For super horizon perturbations, general arguments based on
locality and causality imply that one can use Friedmann equations
with independent constants of motion, within independent Hubble
patches. In the presence of adiabatic perturbations, which are
currently consistent with all cosmological observations (e.g.
\cite{Ade:2013ktc,Ade:2013zuv}), these independent Hubble patches
would only differ in their local value of comoving spatial
curvature ${\cal K}$. This is often quantified using the Bardeen
variable, $\zeta$, where: \beq \delta {\cal K}\equiv \frac{2}{3}
\nabla^2\zeta, \eeq or equivalently the comoving gauge linearized
metric takes the form \beq ds^2= -N^2 dt^2+ a(t)^2
\left[(1+2\zeta) \delta_{ij}+h_{ij}\right] dx^i dx^j,\label{zeta}
\eeq where $h_{ij}$ is a traceless 3-tensor. Planck (+WMAP)
observations \cite{Ade:2013zuv} show that $\zeta$ has a
near-scale-invariant spectrum of perturbations: \beq
\frac{k^3}{2\pi^2} P_{\zeta} (k) = (2.196 \pm 0.059) \times
10^{-9} \left ( k \over 0.05 ~{\rm Mpc}^{-1} \right)^{-0.0397 \pm
0.0073}, \label{p_zeta} \eeq where $k$ is the comoving wavenumber
for spatial fluctuations.

Given that we assumed $Z_2$ (or mirror) boundary conditions for
our 3-brane, we can imagine an atmosphere composed of bulk degrees
of freedom, which is stratified just outside the 3-brane, due to
the gravitational pull of the black hole. Here, we argue that the
thermal fluctuations in the atmosphere of the 3-brane induce a
near-scale-invariant spectrum of curvature perturbations ($\sim$
Eq. \ref{p_zeta}) on our cosmological brane.

Let us first compute the power spectrum of density fluctuations
for a thermal gas of massless scalar particles in
(4+1)-dimensional flat spacetime. The  thermal 2-point correlation
function of a free scalar field is given in terms of the
Bose-Einstein distribution: \beq \langle \varphi(x) \varphi(y)
\rangle_T = \int \frac{d^4k_a}{(2\pi)^4}
\left[\frac{2}{\exp(\omega/T)-1}+1\right]\frac{\exp[i
k_a(x^a-y^a)-i\omega (x^0-y^0)]}{2\omega},\label{2pt} \eeq where
$k_a$ is the spatial wave-number in 4+1D ($ 1\leq a \leq 4$) , and
we used $E=\omega = \sqrt{k^a k_a}$ for massless particles. Now,
using the definition of energy density: \beq \rho(x) = \frac{1}{2}
\dot{\varphi}^2+\frac{1}{2} \partial_a\varphi\partial^a\varphi,
\eeq straightforward manipulations using Eq. (\ref{2pt}) yield
\beq \langle \rho(x) \rho(y) \rangle_T \simeq \frac{5}{8} \left|
\int \frac{d^4k_a}{(2\pi)^4}
\left[\frac{1}{\exp(\omega/T)-1}+\frac{1}{2}\right] \omega  \exp[i
k_a(x^a-y^a)-i\omega (x^0-y^0)]\right|^2.\label{2pt_rho} \eeq

Let us next consider how these density fluctuations affect metric
fluctuations. As a first attempt, we focus on the linear scalar
metric fluctuations in (4+1)-dimensions, which in the longitudinal
gauge can be written as: \beq ds^2= - (1+4\Phi_4) dt^2+(1-2\Phi_4)
\delta_{ab} dx^a dx^b, \label{Phi_4} \eeq where $\Phi_4$ is the
analog of the Newtonian potential.  We can then use 4d Poisson
equation $\nabla_4^2 \Phi_4 \simeq \frac{8\pi G_b}{3} \rho$ to
find the statistics of scalar metric fluctuations. Using Eq.
(\ref{2pt_rho}), we can find the equal-time correlator of
$\Phi_4$: \beq \langle \Phi_4(x^a) \Phi_4(y^a)  \rangle_T \simeq
\frac{5}{8}T^6 \left(8\pi G_b\over 3\right)^2\int \frac{d^4
k}{(2\pi)^4} \frac{ \exp[i k_a (x^a-y^a)]}{k^4}
M\left(\frac{k}{T}\right),\label{2pt_phi4} \eeq where \bea
M(\kappa) &\equiv&  \int \frac{d^4 \kappa'}{(2\pi)^4} \omega_+\omega_- \left[\frac{1}{2}+\frac{1}{\exp(\omega_+)-1}
\right]\left[\frac{1}{2}+\frac{1}{\exp(\omega_-)-1}\right], \\
\omega_\pm &\equiv&
\sqrt{\kappa'^a\kappa'_a+\frac{1}{4}\kappa^a\kappa_a\pm \kappa_a
\kappa'^a}, \eea
while we have dropped the power-law UV-divergent
term ($\propto$ [cut-off]$^6$), e.g. using dimensional
regularization. This UV-divergent term does not depend on
temperature, and presumably can be cancelled with appropriate
counter-terms in other regularization schemes.

Now, we notice that for small $k\ll \Lambda, T$, we have \beq
M(\kappa) \simeq  \frac{15 \zeta_R(5)}{\pi^2} + {\cal O}
(\kappa^2) \simeq 1.576 + {\cal O} (\kappa^2)   ,\label{M_k} \eeq
where $\zeta_R$ is the Riemann zeta function.
Therefore, Eq. (\ref{2pt_phi4}) implies that the 4d Newtonian
potential, due to thermal fluctuations, has a {\it
scale-invariant} power spectrum of the amplitude of $\sim G_b T^3
\sim (T/M_5)^3$.

It is easy to understand this result on dimensional grounds.
Looking at the low frequency limit $\omega \ll T$ of  thermal
density fluctuations (\ref{2pt_rho}), we notice that the argument
inside the absolute value becomes the delta function. In other
words, the densities are only correlated within a thermal
wavelength $T^{-1}$, and only have white noise, or a flat power
spectrum, on large scales\footnote{Note that this is a general
feature of Bose-Einstein distributions, on any space-time
dimension.}.  Then, Poisson equation implies that the potential
power spectrum scales as $k^{-4}$, yielding a logarithmic
real-space correlation function, or equivalently, a flat
dimensionless power spectrum.

So far, all we have done is to study the fluctuations of a
statistically uniform 4 dimensional thermal bath. While the
scale-invariance of this result is suggestive, it is not
immediately clear what this might imply (if anything) for
cosmological curvature perturbations on our 3-brane. To answer
this question, we will first assume that, at some point in its
early cosmological evolution, our 3-brane was in static
equilibrium with its thermal 4d atmosphere. Then a comparison of
(\ref{zeta}) and (\ref{Phi_4}) implies that \beq \zeta(x^i) =
-\Phi_4(x^i,x^4=0) \label{zeta_phi} \eeq
assuming $Z_2$ boundary
conditions at $x^4=0$.  Note that this boundary condition modifies
the thermal spectrum (\ref{2pt_rho}) within a thermal wavelength
of the 3-brane; given that gravity is a long-range force, we do
not expect this to significantly affect the long wave-length
metric fluctuations. Therefore, using (\ref{zeta_phi}), we can put
forth our prediction for the power spectrum of cosmological
curvature fluctuations: \bea
\frac{k^3}{2\pi^2} P_{\zeta} (k) &=& \frac{5}{32\pi^3} \left(8\pi G_b T_b^3\over 3\right)^2 \int_{-\infty}^{\infty}
\frac{dx}{(1+x^2)^2} M\left(\frac{k}{a_bT_b}\sqrt{1+x^2} \right) \nonumber\\
&=& \left[\frac{25\zeta_R(5)}{3072\pi^4} + {\cal O} \left(k \over
a_b T_b \right)^2 \right] \left(T_b \over M_5\right)^6 \nonumber
\\ &\simeq& \left[ 8.66 \times 10^{-5} + {\cal O} \left(k \over
a_b T_b \right)^2 \right] \left(T_b \over M_5\right)^6
\label{final_zeta} \eea
where $T_b$ is the temperature of the bulk
atmosphere, at the moment of equilibrium, where the scale factor
is $a_b$. Furthermore, we used the definition of 5d Planck mass
(\ref{Planck_mass}) to substitute for $G_b$. Comparing Eq.
(\ref{final_zeta}) with Eq. (\ref{p_zeta}) gives the experimental
constraint on the (effective) temperature of the atmosphere: \beq
\frac{T_b}{M_5} = 0.17139 \pm  0.00077, \eeq for the comoving
scale of $k \sim 0.05$ Mpc$^{-1}$. While $T_b$ is for the
atmosphere in the bulk, based on the rate of change in spatial
geometry, we may expect the ``de-Sitter'' temperature of the
boundary to set a minimum for $T_b$. Therefore, we expect: \beq
\frac{H}{2\pi} \lesssim T_b \simeq 0.17 ~M_5, \label{Hubble_M_5}
\eeq

The slight deviation from scale-invariance in Eq. (\ref{p_zeta}),
which is at the level of 4\%, and is now detected with Planck at
$>5\sigma$ level, is not predicted in our simple model of thermal
free 5D field theory. In the next section, we will speculate on
the possible origins of this deviation in our set-up, even though
we postpone a full exploration for future study.

\section{Summary and Discussions}\label{conclude}

In the context of DGP brane-world gravity, we have developed a
novel holographic perspective on cosmological evolution, which can
circumvent a big bang singularity in our past, and produce
scale-invariant primordial curvature perturbations, consistent
with modern cosmological observations. In this paper, we first
provided a pedagogical derivation for the cosmological evolution
of DGP braneworld in FRW symmetry from first principles, and then
connected it to motion in the Schwarszchild bulk geometry,
extending the analyses in \cite{Gregory:2007xy} to realistic
cosmologies. Focusing on the pressure singularity uncovered in
\cite{Gregory:2007xy}, we showed that it is generically
encountered at early times as matter density decays more slowly
than $a^{-4}$. However, we show that the singularity {\it always}
happens  inside a white hole horizon, and {\it only} happens later
than Big Bang Nucleosysntheis (BBN) for a small corner of the
allowed parameter space (i.e. the base of black strip in Fig.
\ref{fig:plot3d}). Therefore, it can never be created through
evolution from smooth initial conditions. This yields an
alternative holographic origin for the big bang, in which our
universe emerges from the collapse a 5D ``star'' into a black
hole, reminiscent of an astrophysical core-collapse supernova
(Fig. \ref{fig:penrose}-left). In this scenario, there is no big
bang singularity in our causal past, and the only singularity is
shielded by a black hole horizon. Surprisingly, we found that a
thermal atmosphere in equilibrium with the brane can lead to
scale-invariant curvature perturbations at the level of
cosmological observations, with little fine-tuning, i.e. if the
temperature is $\sim 20$\% of the 5D Planck mass.

We may go further and argue that other problems in standard
cosmology, traditionally solved by inflation, can also be
addressed in our scenario:
\begin{enumerate}
\item The {\it Horizon Problem}, which refers to the uniform temperature of causally disconnected patches, is addressed, as the ``star'' that collapsed into a 5D black hole could have had plenty of time to reach uniform temperature across its core.
\item The {\it Flatness Problem}, which refers to the surprisingly small spatial  curvature of our universe, is addressed by assuming a large mass/energy for the 5D ``star'', $M_*$.
The radius of the black hole horizon, $r_h$, sets the {\it maximum} spatial Ricci curvature (or minimum radius of curvature) for our universe, and thus can only dominate at late times. If one assumes that the initial Hubble constant is the $\sim$ 5D Planck mass, which is supported by the scale of curvature perturbations above, we have $-\Omega_k \sim (M_5 r_h)^{-2} \sim M_5/M_*$, which could become sufficiently small, for massive stars.

The curvature could of course be detectable at late times, as the Hubble constant drops, depending on the scale of dark energy. However, a detection curvature should generically accompany a detection of large scale anisotropy, as a generic black hole will have a finite angular momentum, which would distort FRW symmetry on the scale of the curvature.
\item The {\it Monopole Problem} refers to the absence of Grand Unified Theory (GUT) monopoles, that should generically form (and over-close the universe) after the GUT phase transition. As we have replaced the singular big bang, with the emergence of a 4D universe at a finite size, the plasma temperature never reaches GUT scale, and thus the GUT phase transition will never have happened in the thermal history of the universe, preventing copious production of monopoles.

To see this, we can translate the observational constraints on the DGP cosmology (normal branch), $r_c \gtrsim 3 H^{-1}_0$ \cite{Azizi:2011ys} into an upper limit on 5D Planck mass:
\beq
H \lesssim M_5 \lesssim \left(H_0 M^2_4\over 6\right)^{1/3}\sim 9~ {\rm MeV} ,
\eeq
where we used the inequality in Eq. (\ref{Hubble_M_5}) to bound the Hubble constant. Correspondingly, the upper limit on the temperature comes from the Friedmann equation in the radiation era, for $g_*$ species:
\beq
T \sim \left(M_4 H\over g_*^2\right)^{1/4} \lesssim 3 \times 10^4 \left(g_* \over 100\right)^{-1/4} ~{\rm TeV}  \ll T_{\rm GUT} \sim 10^{12} ~{\rm TeV}.
\eeq
\end{enumerate}

Yet another attractive feature of our construction is that it
lives in an asymptotically flat space-time. This potentially
allows for an S-matrix description of this cosmology, through
collapse of an ingoing shell,  and emergence of outcoming D-brane.
This might be a promising avenue, especially in light of
significant recent progress in understanding scattering amplitudes
in supergravity (e.g. \cite{Cachazo:2012da,Cachazo:2012pz}.
Furthermore, with trivial modification, this model could also be
embedded in an AdS bulk, which can potentially allow a study of
the strongly-coupled dynamics of emergence through AdS/CFT
correspondence.   Since embedding our braneworld in a large AdS
space-time (instead of Minkowski) simply amounts to adding a small
constant to the right hand side of e.g., Eqs. (\ref{conEn}) or
(\ref{E4}), it need not significantly change any of the
quantitative results that we have found here.

Let us now comment on (some) potential problems.  Perhaps the most
notable problem with the DGP model might be the claim
\cite{Adams:2006sv} that superluminal propagation around
non-trivial backgrounds in DGP model hinders causal evolution, and
UV analyticity/completion. However  this violation of causality is
only a pathology for spacetimes that don't admit a consistent
chronology for (super)luminal signals \cite{Bruneton:2006gf}. Such
spacetimes, e.g. Godel metric, even exist in General Relativity,
and simply point out the absence of {\it global} causal evolution
in those backgrounds. Therefore such geometries cannot emerge out
of classical causal evolution.  The second objection is more
subtle, and relies on the analyticity properties of the scattering
amplitudes for the DGP scalar. However, these conditions (e.g. the
Froissart bound) may be violated in the presence of massless bulk
gravitons. Therefore, these arguments would leave the door open
for a possible UV completion via e.g. string theory and/or AdS/CFT
correspondence.

Another possible pathology of the DGP model is copious spontaneous
production of self-accelerating branes in the bulk
\cite{Gregory:2007xy}, which is estimated via Euclidean instanton
methods. However one may argue that, since self-accelerating
branches have catastrophic ghost instabilities, they should be
excised (or exorcised) from the Hilbert space of the system. Given
that one cannot classically transition from the normal branch to
the self-accelerating branch, this modification would not affect
the semi-classical behavior, but would prevent tunnelling into
unphysical states.

Finally, let us comment on potential testability of this model. As
we pointed out, the simple model of cosmological perturbations,
developed in Sec. \ref{atmosphere} is already ruled out by
cosmological observations at $>5\sigma$ level, as it does not
predict any deviations from scale-invariance. However, it is easy
to imagine small corrections that could lead to a $\sim 4\%$
deviation from scale-invariance, especially given that bulk
temperature is so close (i.e. $\sim 20\%$ of) the 5D Planck
temperature. In the context of our model, the red tilt of the
cosmological power spectrum implies that the amplitude of 5D bulk
graviton propagator, which enters in Eq. (\ref{2pt_phi4}), is
getting stronger in the IR, suggesting gradual unfreezing of
additional polarizations of graviton. For example, this is what
one would expect in cascading gravity \cite{deRham:2007xp}, where
DGP bulk is replaced by a 4-brane, which is itself embedded in a
6D bulk. Similar to the ordinary DGP, the transition in flat space
happens on length-scales larger than $M^3_5/M^4_6$, as the scalar
field associated with the motion of the 4-brane in the 6D bulk
becomes weakly coupled, and boosts the strength of the
gravitational exchange amplitude.

A related issue is that the gravitational Jeans instability of the
thermal atmosphere kicks in for $k < k_J \simeq 0.2\times  T_b~
(T_b/M_5)^{3/2} \sim 10^{-2} T_b $, which may appear to limit the
range of scale-invariant power spectrum to less than the current
observations. However, the time-scale for the Jeans instability
can be significantly longer than the Hubble time, thus limiting
its maximum growth. Nevertheless, one may consider the residual
Jeans instability as a potential origin for the slight red tilt
(i.e. $n_s <1$) of the observed power spectrum. We defer a
consistent inclusion of gravitational backreaction on the 5d
thermal power spectrum (which should account for the impact of
Jeans instability) to a future study.

We should stress that, at this point, the development of a
mechanism responsible for the observed deviation from
scale-invariance is the most immediate phenomenological challenge
for our scenario. The next challenge would be a study of the
interactions that lead to deviations from scale-invariance, and
whether they satisfy the stringent observational bounds on
primordial non-gaussianity \cite{Ade:2013ydc}. Other interesting
questions might be, given that the emergence from the 5D  black
hole might happen at relatively low temperatures, could there be
observable predictions for gravitational waves (either on
cosmological scales, or for gravitational wave interferometers),
or even modifications of light element abundances in Big Bang
Nucleosynthesis.

Ultimately, an entire new world might emerge ``Out of the White
Hole'', and replace Big Bang with a mere mirage of a non-existent
past!
\bigskip

{\it Acknowledgements} We would like to thank Nima Doroud, Gregory Gabadadze, Ruth
Gregory, Rob Myers, Paul McFadden, Eric Poisson, Kostas Skenderis,
Misha Smolkin, Marika Taylor and Herman Verlinde for inspiring
discussions and useful comments. This work was supported by the
Natural Science and Engineering Research Council of Canada, the
University of Waterloo and by Perimeter Institute for Theoretical
Physics. Research at Perimeter Institute is supported by the
Government of Canada through Industry Canada and by the Province
of Ontario through the Ministry of Research \& Innovation.

\bibliography{FRW}{}
\bibliographystyle{JHEP}

%
%
%
%

\end{document}